# Optical projection and spatial separation of spin entangled triplet-pairs from the $S_1$ ($2^1A_g^-$) state of pi-conjugated systems

[1]Raj Pandya, [1]Qifei Gu, [1]Alexandre Cheminal, [1]Richard Y. S. Chen, [1]Edward P. Booker, [2]Richard Soucek, [2]Michel Schott, [2]Laurent Legrand, [3]Fabrice Mathevet, [1]Neil C. Greenham, [2]Thierry Barisien, [4]Andrew J. Musser*, [2]Alex W. Chin* and [1]Akshay Rao*

* correspondence: ajm557@cornell.edu, alex.chin@insp.jussieu.fr, ar525@cam.ac.uk

[1]Cavendish Laboratory, University of Cambridge, J.J. Thomson Avenue, CB3 0HE, Cambridge, United Kingdom

[2] Sorbonne Université, CNRS, Institut des NanoSciences de Paris, INSP, 4 place Jussieu, F-75005 Paris, France

[3]Sorbonne Université, CNRS, Institut Parisien de Chimie Moléculaire (IPCM) UMR 8232, Chimie des Polymères, 4 Place Jussieu, 75005 Paris, France

[4]Department of Chemistry and Chemical Biology, Cornell University, Baker Laboratory, Ithaca, NY, 14853, U.S.A

Abstract

The $S_1$ ($2^1A_g^-$) state is an optically dark state of natural and synthetic pi-conjugated materials that can play a critical role in optoelectronic processes such as, energy harvesting, photoprotection and singlet fission. Despite this widespread importance, direct experimental characterisations of the electronic structure of the $S_1$ ($2^1A_g^-$) wavefunction have remained scarce and uncertain, although advanced theory predicts it to have a rich multi-excitonic character. Here, studying an archetypal polymer, polydiacetylene, and carotenoids, we experimentally confirm that $S_1$ ($2^1A_g^-$) is a superposition state with strong contributions from spin-entangled pairs of triplet excitons ($^1$(TT)). We then more importantly show that optical manipulation of the $S_1$ ($2^1A_g^-$) wavefunction using triplet absorption transitions allows selective projection of the $^1$(TT) component into a manifold of spatially separated triplet-pairs with lifetimes enhanced by up to one order of magnitude. Our results provide a unified picture of $2^1A_g^-$ states in pi-conjugated materials and show that the stimulation of the bound $^1$(TT) optical absorption offers a hitherto unexplored pathway to create additional near-free triplets. More generally, our findings open new routes to exploit $2^1A_g^-$ dynamics in singlet fission, photobiology and for the generation of entangled (spin-1) particles for molecular quantum technologies.

Introduction

Electronically conjugated polymers and oligomers are ubiquitous in biological systems, with nature deploying these flexible and chemically tunable systems for a wide variety of advanced optoelectronic functions[1–3]. For many photosynthetic organisms they play a vital dual role as both light harvesting antennae and photoprotective molecules that can remove deleterious excess excitations[2,4,5]. In synthetic molecular materials developed for organic electronics they have transformed transistor and light-emitting diode technology[6–8], as well as becoming promising components for next-generation photovoltaic (PV) devices with the potential to overcome the Shockley–Queisser limit *via* singlet fission (SF). In SF the absorption of a single photon ideally results in the formation of two spatially separated triplet excitons[9–12], although a concerted sequence of ultrafast electronic and vibrational dynamics must compete with both radiative and non-radiative losses for this exciton multiplication to be efficient enough for applications[13–17]. Observation, understanding and control of these many-body quantum dynamics is therefore essential for optimisation of conjugated molecules for technologies such as SF-enhanced PVs[18–21].

However, the nature and understanding of electronic excitations in quasi-1D pi-conjugated polymers is greatly complicated by large system sizes, low dimensionality and very strong Coulombic interactions[22–30]. This leads to pronounced and well-studied effects of electronic correlation in these systems, with the key result being that their optical properties are dominated by strongly bound Frenkel excitons (electron-hole pairs) with very large (0.1-1 eV) exchange splittings between the optically excitable singlet excitons and the lowest (dark) triplet excitons. As a result of the low energy of the triplet exciton state ($T_1$), creating a pair of triplet excitations may in fact cost less energy than the excitation of one singlet exciton. In this case – and as two spin-1 particles can possess a total spin of 0, 1, or 2 – it then becomes possible that the lowest-lying singlet excited state ($S_1$) could be formed from pairs of triplets excitons with zero net spin, *i.e.* a singlet state. We note here that for a pair of triplet excitations to have zero total spin requires strong quantum correlations between the individual spins of each exciton. Indeed, in the language of quantum information theory, the singlet spin wavefunction for two spin-1 particles is said to be a Bell, or maximally entangled state[31]: if the individual particles could be spatially separated and subjected to independent spin measurements, the results of these measurements would be predicted to violate Bell's inequalities. We shall use the term entanglement henceforth to refer to the potentially long-range (see below) spin correlations between triplet excitons in the materials that we shall study[32].

In this work we focus on linear pi-conjugated systems with $C_{2h}$ symmetry, which allows us to label the electronic states of the system according to how their wavefunctions transform under the point group symmetry operations. The ground state ($S_0$) is always of $A_g^-$ symmetry, so is denoted as $S_0$ ($1^1A_g^-$), and the lowest triplet state is $T_1$ ($1^3B_u^+$). Consequently, the product wavefunction of a pair of triplets has an overall $A_g$ symmetry and, as discussed above, it is then possible for $S_1$ to have the same symmetry as the ground state, where it is denoted as the $S_1$ ($2^1A_g^-$) state[22,33–37]. The second excited state ($S_2$) is typically of $1^1B_u^+$ symmetry and one-photon excitation predominantly occurs to this state[38,39]. We note that triplet-pairs are not the only electronic configurations that could contribute to $S_1$ ($2^1A_g^-$), and some other singly excited configurations are shown in Figure 1. Indeed, what is referred to as the $S_1$ ($2^1A_g^-$) state is, in general, a complex superposition of single, double and higher-order electronic configurations of overall $A_g$ symmetry and zero spin[39–41].

Electron correlations within $S_1$ ($2^1A_g^-$) bring its energy below that of $S_2$ ($1^1B_u^+$) for sufficiently long polyenes (molecules containing a minimum of three alternating double and single carbon–carbon bonds), with profound consequences. Following one-photon excitation to $S_2$ there is usually rapid and efficient internal conversion to $S_1$; the shared symmetry of $S_0$ and $S_1$ then means that systems become non-fluorescent[39,40,42–46]. Once formed, these dark states then internally convert (IC) to the ground state, form triplet excitons by intersystem crossing (ISC), or – in the case of shorter polyenes – may lead to photoisomerisation events[41,43]. In smaller systems, the former and latter processes are dominant due to the weak spin-orbit coupling for ISC and much stronger vibronic coupling for IC and structural relaxation. However, extended polymers can be large enough to accommodate spatially well-separated and essentially non-interacting triplet excitations, raising the intriguing possibility of intramolecular-SF from the bound 'triplet-pair' excitations that formally contribute to the $S_1$ ($2^1A_g^-$) wavefunction (see Figure 1b). Moreover, as rapid and therefore efficient triplet production can only occur *via* spin-allowed (spin conserving) processes, free triplets formed by intramolecular-SF must be created in entangled singlet spin states.

Various strategies have been suggested for biasing the fate of $S_1$ ($2^1A_g^-$) towards triplet production over internal conversion, such as engineering an enhanced 'triplet-pair' weight into the $S_1$ ($2^1A_g^-$) wavefunction with donor-acceptor copolymers[26], and exploiting polymer-polymer interactions to extract long-lasting triplet states[47,48]. Interestingly, when singlet fission does occur in systems expected to have relevant $S_1$ ($2^1A_g^-$) states, the optical signatures of this intervening singlet are not observed[18,49–53], and its proposed role in fission remains controversial[26,39,54,55]. Although, several groups have shown that some polymers, all the way back to polyacetylene[39,40] have triplet-pair-like states similar to $2^1A_g^-$ [49,56], the general design strategy has been to avoid 'dark' $2^1A_g^-$ states for efficient fission[18]. The biological implications of both intramolecular-SF and the $S_1$ ($2^1A_g^-$) state in carotenoids have also been discussed in the context of photosynthetic light-harvesting proteins[43,57–60]. Recent theoretical works have shown in these biosystems that in specific geometries the $2^1A_g^-$ wavefunction can have significant admixtures from other states and is not a pure $^1(TT)$ excitation[61]. Yet much about the electronic structure and ultrafast dynamics of $S_1$ ($2^1A_g^-$) excited states remains unknown.

Unfortunately, probing real-time fission from $S_1$ ($2^1A_g^-$) states is generally complicated by both fundamental and practical factors. Firstly, the $S_1$ ($2^1A_g^-$) state is dark and often decays through non-radiative pathways in just a few picoseconds. Secondly, many theoretical properties of correlated 1D electronic systems emerge in the *idealised* polymer limit of very large and perfectly ordered system sizes[22,33,34,36,37,39], which in most experimental polymer systems are difficult to realise due to intra-chain disorder, cross-linking and integration into nanostructures such as proteins.

Here, we overcome these limitations by applying ultrafast nonlinear optical techniques to *manipulate* the $S_1$ ($2^1A_g^-$) state during its lifetime, thereby altering its dynamical fate in two classes of organic material: (1) disorder-free topochemical polydiacetylene crystals and (2) carotenoids in isolated and aggregated forms. Strikingly, we show that optical pulses centred on well-known triplet-pair-derived absorption transitions can effectively project the $S_1$ ($2^1A_g^-$) wavefunction into a manifold of spatially separated, quasi-free triplet-pair states. The production of geminate, spin-entangled triplet-pairs is evidenced by their motion-limited recombination to the ground state, giving us additional insight into triplet diffusion and re-binding during relaxation back to $2^1A_g^-$ (Figure 1b). Performing experiments on 'twisted' polydiacetyelenes where the $2^1A_g^-$ and $1^1B_u^+$ energies are reversed, *i.e.* a bright $S_1$ state, varying solvent, defect density, sample temperature, *etc*, and studying carotenoid monomers of differing conjugation length, allows the electronic signatures observed to be robustly assigned. Tuning the energy and temporal overlap of optical pulses with the various electronic states also enables a microscopic mechanism for triplet-pair liberation from $2^1A_g^-$ to be proposed. This involves coulombic repulsion between localised and delocalised electron-hole pairs within a triplet-pair state. The results have implications for the design of new conjugated molecules for intramolecular-SF, photoreception and photoprotection, and show that the $2^1A_g^-$ state can in-fact be a useful pathway for near-free triplet generation, by the hitherto unexplored mechanism of stimulation of the bound $^1(TT)$ optical absorptions. More generally the results highlight the prospect for exploiting and controlling entangled states in organic systems, providing the possibility for room temperature quantum circuitry in molecular materials[62].

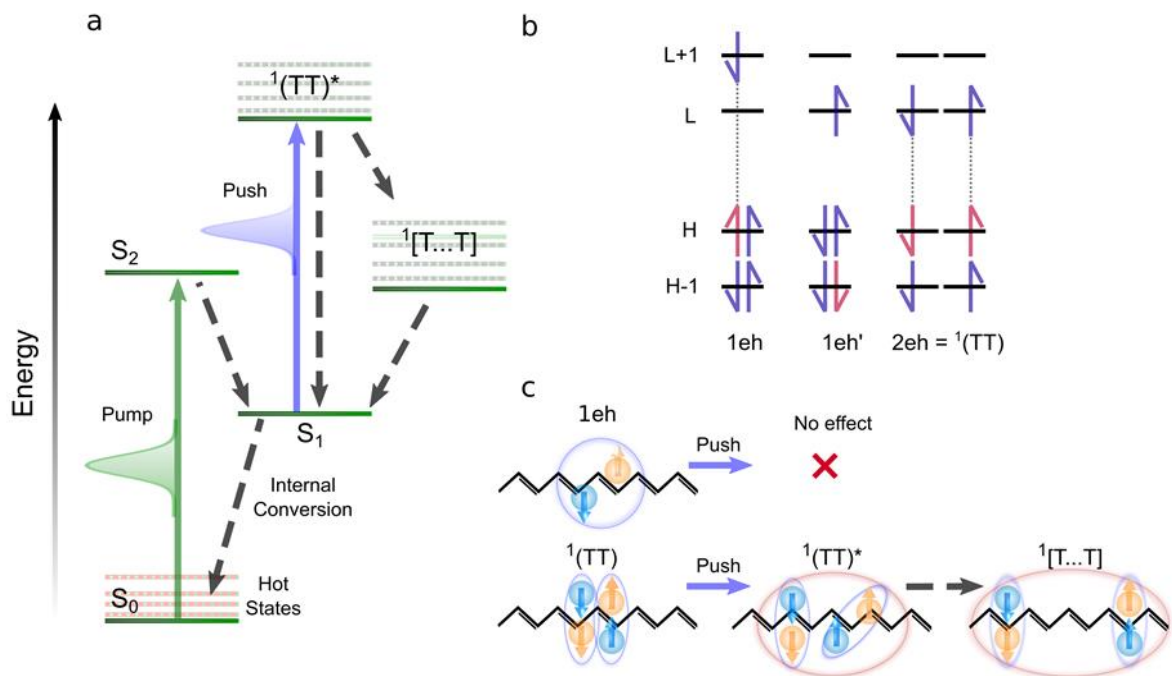

**Figure 1: Energy levels, dynamics and electronic properties of the key excited states of 1D pi-conjugated polymers of $C_{2h}$ molecular symmetry. a.** Schematic of energy levels and dynamics measured in the pump-push-probe experiments detailed in the main text (pump – green arrow; push – blue arrow). Because of the $^1(TT)$ amplitudes in the $S_1$ ($2^1A_g^-$) wavefunction, this state can be tracked and selectively projected into the triplet-pair manifold by pushing the excited-state absorption transition $^1(TT)$ – $^1(TT)^*$. The spatially separated triplet-pairs ($^1[T...T]$) appear from relaxation of $^1(TT)^*$ (a $B_u$ symmetry state). **b.** Some possible electronic configurations of the frontier molecular orbitals that are superposed together by the configuration interaction into the low lying $S_1$ ($2^1A_g^-$) wavefunction. Red arrows indicate 'hole' excitations, electrons are shown in blue. 1eh indicates an electronic configuration containing one electron-hole pair in an excitonic spin-singlet state. The singlet $^1(TT)$ configuration is a doubly excited configuration that can be seen as a pair of low energy triplet excitons with anti-aligned spins. **c.** Illustrative sketch of some possible spatial wavefunctions of electrons and holes for the configurations shown in **b**. The transition we stimulate with the push pulse is derived from the transition of single, isolated triplets, *i.e.* it represents an excitation of a single triplet component within the $2^1A_g^-$/$^1(TT)$ state (as opposed to an excitation of the entire multiexciton into a putative T*T* configuration). Furthermore, the push chiefly acts on the $^1(TT)$ part of the wavefunction (bottom) and not the 1eh (or 1eh′) amplitudes (top). A mechanism is suggested for the push-induced separation of the triplet-pairs from $^1(TT)^*$, whereby one of the excited triplets is highly delocalised, promoting relaxation channels that separate the triplets far enough apart to behave as near-free particles. We note that the triplet spins remain correlated (red circles) in this process, as described in the main text.

Topochemically polymerised polydiaceytlene (PDA) chains can take two different geometrical conformations, nominally referred to as the 'red' and 'blue' phases. In the former case the $2^1A_g^-$ state lies above a 'bright' state of $1^1B_u^+$ symmetry, whereas for the 'blue' phase this arrangement is reversed and chains are non-luminescent. In this work we deal exclusively with the more correlated and SF active[26,55,63] 'blue' chains. Uniquely we study chains a few micrometres long, embedded in a crystal of their diacetylene monomer, highly ordered and isolated from one another (~100 nm)[64]. Figure 2b shows the absorption spectrum of a PDA crystal with light polarised parallel, perpendicular and at 27.5° with respect to the long axis of the chain. When light is polarised parallel to the double and triple bond axis, the absorption is maximum with a strong zero-phonon line observed around ~635 nm (1.95 eV), and a vibronic progression out to 530 nm (2.35 eV)[64]. On rotating the polarisation of light, the absorbance is reduced to zero within the resolution of our instrument, evidencing a near perfect orientation of chains (see X-ray diffraction; SI, Figure S1). This near absence of structural disorder extends to the electronic properties highlighted by the high oscillator strength (~$10^6$ cm$^{-1}$), strong coupling between the monomer units and macroscopic exciton coherence lengths[65,66]. PDA can therefore be considered as an ideal model system to investigate the excited state properties of pi-conjugated polymers and the links between SF and correlated electronic excitations across nm-µm and fs-ps scales.

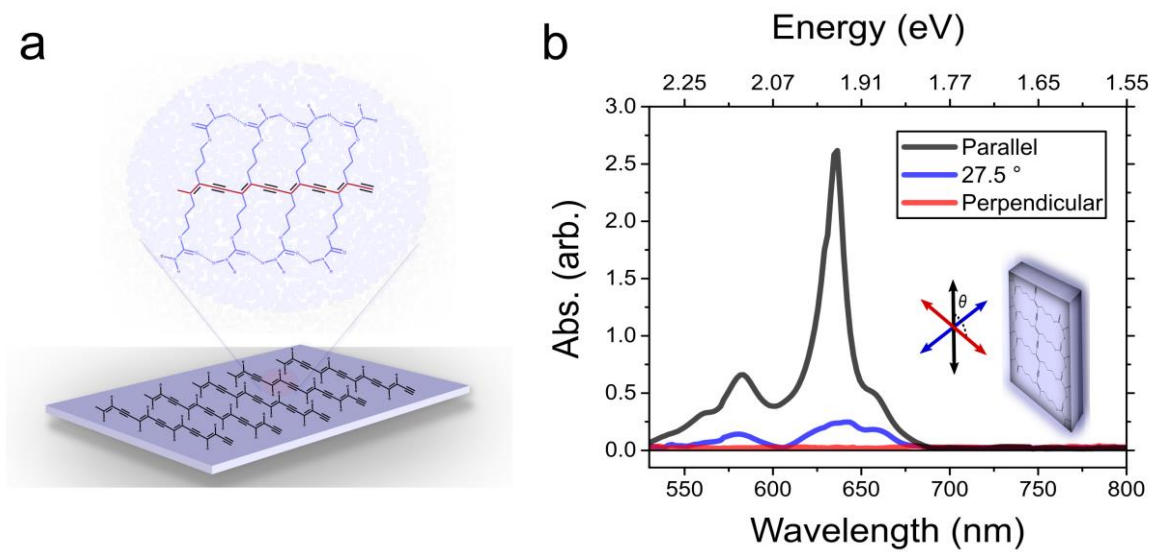

**Figure 2: Optical characterization of polydiacetylene. a.** Cartoon of 3-BCMU PDA crystal showing long, highly-ordered polymer chains; the chains lie near perfectly orientated in the experimental crystal as depicted in the cartoon (3-BCMU refers to the –$CH_2COOC_4H_9$; stabilising group). **b.** Absorption spectra of aligned PDA chains within a dilute matrix of diacteylene monomer, with light polarised parallel, at 27.5• and perpendicular to the long axis of chains. A strong zero-phonon line is observed at 635 nm, with vibronic peaks arising from the double and triple bonds at ~575 nm and ~555 nm respectively. An extra shoulder at the band edge (665 nm) arises from partially polymerised chains.

To initially establish the correlated triplet character of the $2^1A_g^-$ state in PDA, we first carried out two pulse broadband pump-probe experiments with an ultrashort 10 fs pump pulse tuned to overlap with the excitonic transitions of 'blue' PDA (550 – 650 nm). Here, we spectrally resolve the change in probe transmission, $\Delta T$, with and without the pump pulse incident on the sample and normalise by the probe-only transmission, $T$ to yield a $\Delta T/T$ signal. Positive signatures in pump-probe correspond to a ground state bleach (GSB) or stimulated emission (SE), whereas negative features are indicative of a photoinduced absorption (PIA).

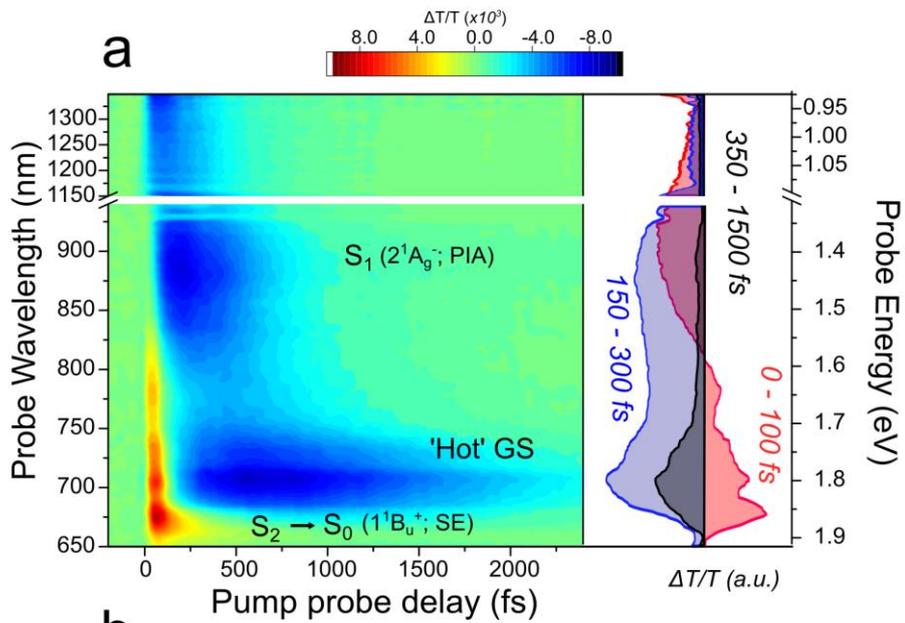

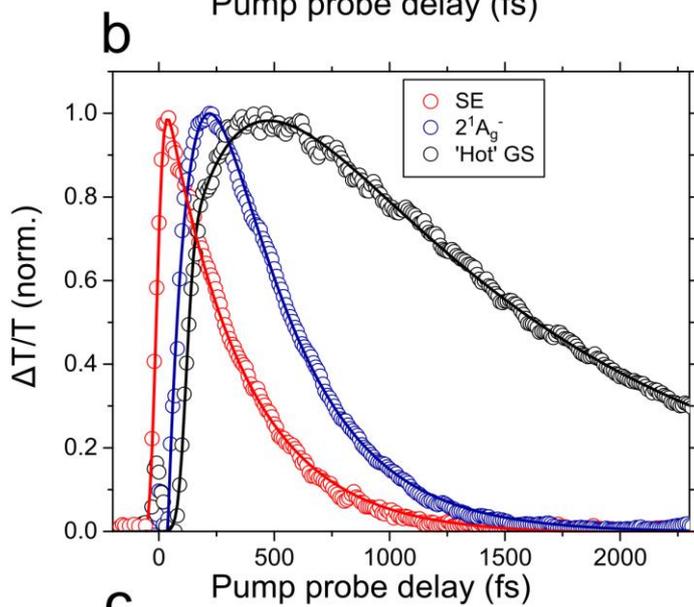

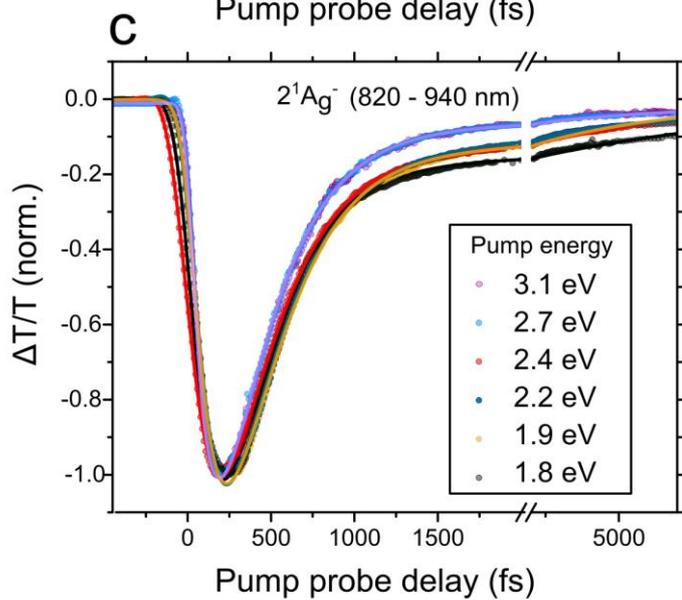

**Figure 3: Transient electronic response of 3-BCMU polydiacetylene following photoexcitation. a.** Femtosecond pump-probe spectrum of 'blue' PDA following photoexcitation with a 10 fs pulse centred at 530 nm (FWHM ~65nm). Three distinct regions can be observed: a sharp positive band centred around ~650 nm corresponding to the stimulated emission (SE), the photoinduced absorption (PIA) of the $2^1A_g^-$ state at ~870 nm and a negative band at ~750 nm resulting from the vibrationally 'hot' ground state ('Hot' GS). The integrated spectra are shown alongside. **b.** Kinetics (circles) and fits (line) associated with stimulated emission, 'hot' ground state and $2^1A_g^-$ state. The data is obtained following spectral decomposition of the pump-probe response with a genetic algorithm and is normalized for ease of comparison (see SI, Figure S2 for spectra). **c.** Dependence of the dynamics of the $2^1A_g^-$ state on pump energy (1.8 – 3.1 eV, ~200 fs pulse, FWHM ~10 nm). The dynamics are effectively pump energy independent (constant excitation density) with a small variation observed when pumping close to the band edge and at high energies.

**Pump-Probe.** In Figure 3a we present the pump-probe spectra from our two-pulse pump-probe experiments; unless otherwise stated all experiments are carried with pump, probe (and push) parallel to the long axis of the PDA chains. Following spectral decomposition using a genetic algorithm (see SI, Figure S2), we identify three distinct species in the spectra whose decays can all be fitted with a single exponential (Figure 3b). Between 650 – 800 nm, we observe a rapid (~320 ± 5 fs) decay of the PDA SE from the $1^1B_u^+$ ($S_2$) state. The SE consists of a vibrational progression of peaks which are slightly red shifted compared to that previously reported[67–69] due to some small reabsorption in the crystal; measurements on thinner (~100 nm thick) crystals indicate reabsorption effects play no role in the dynamics reported throughout this study (see SI, Figure S3). Additionally, between 820 – 1350 nm there is an energetically broad band that has been previously assigned to the $2^1A_g^-$ photoinduced absorption (PIA)[70]. Vibronic relaxation from $1^1B_u^+$ to $2^1A_g^-$ means the growth of the $2^1A_g^-$ PIA is delayed with respect to that of the SE; $2^1A_g^-$ then decays with an exponential time constant of ~380 ± 5 fs. The results above are fully consistent with previous pump-probe data and state assignments on a range of 'blue' PDAs[67,68,71].

However, we also note a prominent third species between 675 – 750 nm which grows from the $1^1B_u^+$ and $2^1A_g^-$ states ($\tau_{rise}$ ~350 ± 10 fs). To the best of our knowledge this latter PIA has not previously been observed or assigned in the spectra of PDA. Comparable features are known in other systems with rapid non-radiative decay such as the polyenes, and they are linked to the deposition of significant electronic energy into highly excited ground-state vibrational modes[51,72–74]. The resulting vibrationally 'hot' ground state is somewhat closer in energy to the optically allowed state and consequently exhibits red-shifted absorption. Given the energy of the PIA is higher than that reported or expected for any $^1$(TT) or free triplet transition in PDA[49,68], with the lifetime being also too short, we assign it to a 'hot' ground state (further discussion SI, Figures S4-S7).

Returning to the decay of the PIA associated with the $2^1A_g^-$ state, we find it is largely pump excitation energy and temperature independent (Figure 3c; similar behaviour at 6 K, see SI, Figure S4). These results demonstrate that simply providing the system with increasing amounts of energy does not prolong the $2^1A_g^-$ lifetime or significantly change the electronic dynamics, implying rapid dissipation of excess excitation energy. The invariance of the dynamics with pump energy is a crucial point for our pump-push-probe experiments, where the total combined energy provided by pump and push falls within the range of one-photon excitation energies considered in Figure 3c.

**Pump-push-probe** To directly interrogate and optically control the population of the $2^1A_g^-$ state we extend the pump-probe experiment using a third 'push' pulse[72,75,76] which acts on the photoexcited chains. Here, the same ultrashort pump pulse (10 fs, 550 - 650 nm) photoexcites the PDA chains and is subsequently followed by a 200-fs push pulse at a fixed time delay. We emphasise that the excess energy in the pump with respect to the ~635 nm bandgap appears to have no influence on the observed $\Delta T/T$ signal (see Figure 3c and SI, S7). The push pulse is tuned to be resonant with the excited state absorption of the $2^1A_g^-$ state and promotes a sub-population to higher lying excited states. The resulting signals are then measured in the same way as for pump-probe by a delayed probe pulse. Comparing push 'on' and 'off' signals allows us to extract the push-induced change in differential transmission, $\Delta(\Delta T/T)$. Using the pulse sequence as shown in Figure 4a enables us to measure the pump-probe, push-probe and pump-push-probe signals within the same experiment. Recording the push-probe signal separately allows us to ensure that the push fluence is kept sufficiently low, preventing excitation of the ground state *via* multi-photon absorption. In the pump-push-probe of PDA, negative signals indicate a push-induced increase in the population of a given pump-excited electronic state; positive signals conversely correspond to a depletion of the excited state population by the push pulse.

Figure 4b shows the $\Delta(\Delta T/T)$ response of PDA chains, with the push pulse tuned to 940 nm and delayed to 400 fs after the pump, such that it arrives immediately following the maximum in $2^1A_g^-$ population (qualitatively similar results are obtained using pump-push delays of 200 fs and 800 fs in the rise and decay of $2^1A_g^-$ SI, Figure S9). We observe two negative bands, one centred at 700 nm in the spectral region corresponding to the 'hot' ground state and a second at 870 nm, in line with $2^1A_g^-$ excited state absorption. The $\Delta(\Delta T/T)$ kinetic of the 'hot' ground state decays at a rate similar to that observed in our pump-probe experiments ($\tau \sim 2.3 \pm 0.05$ ps). Conversely for the $2^1A_g^-$ state, the push pulse suppresses the excited state decay, increasing the lifetime of the population in the spectral region associated with the $2^1A_g^-$ PIA by one order of magnitude from ~0.38 ps to ~15 ps (Figure 4c,d). Furthermore, although the exact rise time of the push-induced kinetic cannot be resolved in these experiments (IRF ~200 fs), examining the spectra in Figure 4e shows that the $2^1A_g^-$ and 'hot' ground state PIAs are present almost instantaneously following the push pulse; this is in stark contrast to the pump-probe spectra where $2^1A_g^-$ and 'hot' ground state PIAs develop sequentially. The pump-energy dependent measurements (Figure 3c) show that simply introducing additional energy into the system does not prolong the lifetime of the transitions, hence our observations reflect something other than simple deposition into the electronic state of an additional ~1.3 eV by the push. Additionally, the spectrum of the pushed $2^1A_g^-$ state narrows from ~110 meV to ~60 meV between 700 fs and 7 ps (Figure 4f), as compared to a final linewidth of ~95 meV for the unpushed 'cold' $2^1A_g^-$ state signature.

We can begin to rationalize these results by first noting that previous studies have identified a strong pump-probe PIA signal at ~1.36 eV (910 nm) that has been unambiguously assigned to excited state absorption from the lowest triplet state[68,70,77,78]. The higher-lying excited triplet state is denoted as T* and has an experimentally determined energy in the region of 2.3-2.4 eV[67,68]. With experimental works placing the energy of the lowest triplet state at around 0.9-1.0 eV[68,70,77], the energy for creating a free triplet in the lowest state and a free triplet in the T* state from $S_1$ ($2^1A_g^-$) is 1.35-1.55 eV [68,78], which matches near perfectly with the position of the $2^1A_g^-$ PIA we observe. We therefore consider the possibility that the existence of triplet-pair amplitudes in the $2^1A_g^-$ wavefunction (see Figure 1) allow $S_1$ to be projected into an excited manifold consisting of doubly excited configurations involving the coexistence of a low energy triplet and an optically excited triplet. However, due to Coulombic interactions, there will generally be a broad range of physical eigenstates corresponding to such triplet pairs which we denote as $^1(TT)^*$ ($B_u$ symmetry in $C_{2h}$ space group). These states, all of different spatial triplet separations, structure geometries and internal kinetic energies, contribute to the overall broadness of the PIA feature, as does the ultrafast relaxation of the $2^1A_g^-$ state itself. It is important to note that our push pulse does not need to excite the $T_1$-T* transition of an individual triplet, but rather bring $^1(TT)$ into a higher $^1(TT)^*$ level. The energy for this will be similar, but not identical, to the excitation of a single triplet into T*[79]. Hence although the 940 nm push pulse used does not precisely match the $2^1A_g^-$ – T + T* energy gap, it is sufficiently resonant with the lower-energy 'tail' of the (cooled) broad $2^1A_g^-$ PIA to push triplet-pair amplitudes in the superposition to higher $^1(TT)^*$ levels. Even more strikingly, we now demonstrate how this picture allows us to understand the lifetime enhancement and narrowing of the $2^1A_g^-$ PIA as a result of *near-free* triplet ($^1[T…T]$) generation from $2^1A_g^-$ and the subsequent motion of well-separated and maximally spin-entangled excitons under 1D (intramolecular) quantum confinement.

Following the push there will be an instantaneous transfer of population from $2^1A_g^-$ to $^1(TT)^*$, Figure 1a. Some of this population rapidly (sub-200 fs; Kraabel *et al.* measured a T* lifetime of 100 fs[68]) relaxes back to $2^1A_g^-$, releasing energy in the form of high energy phonons (heat) to the surrounding monomer units. This results in population of vibrationally excited levels of $1^1A_g^-$ leading to the instantaneous push-induced 'hot' ground state signature observed at 700 nm. However, some of the population pushed into $^1(TT)^*$ decays *via* a separate pathway. Based on the enhanced lifetime and narrow line shape of the signature at ~870 nm in Figure 4b, we suggest that some of the population in $^1(TT)^*$ can access a different part of the potential energy surface and form weakly bound triplet-pairs ($^1[T…T]$), in which the individual triplets are spatially separated but still spin entangled. Because freer triplet-pairs will have a have narrower density of states, their PIA bands are expected to be sharper[68,80,81]. Given the line-narrowing shown in Figure 4f is greater than the original cold $2^1A_g^-$ PIA (*i.e.* at 1 ps in Figure 3a), we suggest this to be associated with the freer triplet-pair state as opposed to vibrational cooling. This is further supported by the fact that the long-time species in Figure 4f is slightly red-shifted from that on early timescales, whereas vibrational cooling would give rise to a blue-shift; we note several studies other studies have assigned such observations accordingly[80–82]. Any excess energy, which likely plays a role in initially separating the bound triplet-pairs, may be rapidly (~200-500 fs) dissipated following the push pulse. These 'near-free' triplet states do not overlap and cannot internally convert as effectively as they do when 'bound' in the $2^1A_g^-$ state, so decay more slowly. Consequently, the pump-probe kinetic with the push 'on' at ~870 nm will have a longer decay as compared to that with push 'off'. When the two are subtracted this will result in a $\Delta(\Delta T/T)$ signal with both an enhanced lifetime and rise. The latter will be a convolution of several processes including the time taken to transition between the $^1(TT)^*$ and $^1[T…T]$ states and the underlying decay of $2^1A_g^-$, and as a result the details of the transition from the $^1(TT)^*$ to $^1[T…T]$ manifold is challenging to ascertain from this data alone. Eventually the spatially separated triplet-pairs diffuse and recombine resulting in the overall decay of the $\Delta(\Delta T/T)$ kinetic. Given that no long-lived (μs) signal corresponding to isolated triplet-transitions is observed, the recombination must involve triplet-pairs being annihilated. As we only measure recombination of geminate pairs to the singlet ground state in PDA, the enhanced lifetime we measure implies that the decorrelation time of the triplet spins must be longer than 15 ps at room temperature: any spin-1 pairs would quickly annihilate to leave one triplet in the long-lived $T_1$ state. The absence of additional push induced signals suggests $^1[T…T]$ generation is relatively efficient; this is in-line with the strong electronic correlations and absence of numerous 'dark'/CT branching states in 'blue' PDA[64]. Given that ISC and the intrinsic spin evolution of geminate SF triplets occurs on nanoseconds time scales in other organic SF systems[82], we therefore think it highly likely that the triplets remain spin-entangled over the separation, diffusion and recombination process. We note that we have interpreted our results in terms of the single- and two-particle states shown in Figure 1b,c that constitute the primary components of the $2^1A_g^-$ and $^1[T…T]$ wavefunctions. We recognise that this is a simplification of the complex structure of the states in these highly correlated systems. However, a fully detailed picture of the dynamics of highly excited states in these systems is beyond the scope of current methods, and our independent-particle approach allows for intuitive connections to other materials.

It is important to consider other possible origins of the observed signals. Electronic disorder in the PDA crystals used here (0.5 mm thickness) is small as evidenced by the low Urbach energy ($E_u$ ~31 meV; SI Figure S3). Thinner crystals (100-200 nm) have a greater energetic disorder ($E_u$ ~38 meV). If defects were responsible for the observed push lifetime enhancement and spectral narrowing between 820-920 nm, one might expect in a more defective crystal the effect to be stronger. This might be *via* new signatures or a longer decay time. Both 0.5 mm thick and 100-200 nm thick PDA crystals show identical pump-probe and pump-push-probe responses, suggesting structural or electronic disorder, which contributes to the Urbach energy, is not to be responsible for the observed lifetime enhancement. Another possibility could be that the push pulse generates free charge in PDA. In this case a new photoinduced absorption band corresponding to free charges would be expected. Comeretto *et al.* investigated the energy and dynamics of free charge PIAs in PDA and found two bands sitting at 0.82 eV and 0.95 eV respectively with millisecond lifetimes[83,84]. The difference in spectral position and decay kinetics between these bands and the response observed here further rules out the possibility of free charge creation by the push pulse. One final possible consideration is that the push activates dark, intermediate charge-transfer or polaron states. In measurements of $C_{60}$ doped and carbazolyl modified PDAs[85,86], no intermolecular charge-transfer bands were observed, suggesting CT exciton formation to be unfavourable. This may also explain why primarily only the triplet-pair amplitudes in the $2^1A_g^-$ wavefunction appear to be addressed by the push. Inter-chain CT states have been observed in PDA, but only in crystals with high polymer contents ($10^{-2}$ by weight), well above those studied here[87]. Furthermore, the PIA associated with an *intramolecular* CT state (transitioning from $2^1A_g^-$) would be expected to be at ~0.9 eV [88,89]. At this push energy however, we observe no $^1[T…T]$ signatures, bolstering the claim that the push selectively addresses triplet-pair amplitudes in $2^1A_g^-$. The 'mirror image' between absorption and emission spectra as well as high electron mobility also suggests singlet-polaron formation to be disfavoured[90]. Furthermore, in pump-probe measurements, stimulated emission from the zero-phonon band does not spectrally shift (within our resolution of ~1.2 nm) following photoexcitation (SI, Figure S4). Further supporting the idea that there is very little change in the lattice between ground and excited state in PDA, with singlet polaron formation generally disfavoured. We note there is a 10 meV shift in the SE from the first vibronic peak, 500 fs following photoexcitation; this is likely as a result of reabsorption. Triplet-polarons have been evidenced in 'blue' PDA, but only under intense UV irradiation, well above the ~3.5 eV total pump and push energies used here[91]. Finally, three, largely uncharacterised, dark states have been reported in 'blue' PDA termed $X_{1/2}$ and Y. The PIAs associated with these states lie between 1.75 to 2 eV and given their lifetime is ~200 fs we do not consider them to be of significance for the results presented here[67]. In previous studies, of PDAs fission has been observed, with a maximum yield in the 'blue' PDA of <1 % [67,68] with pulse intensities four times higher than the maximum fluences explored in Figures 3 and 4. For our work, where we focus on the main 'singlet' transition, even in conditions giving maximum fission yield, 'free' triplets should not make a significant contribution to the signals detected. The 'intrinsic' triplet formation is of such low efficiency, that free triplet formation/annihilation is negligible (and not detected) in the regime of pump intensities we focus on. We note that similar effects have been observed in carotenoids by multiphoton excitation from the ground-state[92,93]; a process with inherently different selection rules to that studied here (see later). Ruling out a wide range of other possible explanations further strengthens our assignment of the push-induced signature being from a $^1$[T...T] state.

Polymer chain geometry plays an important role in the electronic structure of PDAs. In the case of 'blue' PDA, chains are planar and the $1^1B_u^+$ state lies above $2^1A_g^-$. However, on twisting of the polymer backbone, such that there is a 14° torsional angle, the energy of $2^1A_g^-$ rises above $1^1B_u^+$, comparable to the behaviour observed in other polyenes with respect to conjugation length[42]. The material then becomes luminescent and is called 'twisted' PDA (Figure 5a)[94]. Repeating pump-probe experiments on 'twisted' PDA shows a broad PIA (decay time ~13.8 ps) lying between 750 to 900 nm, similar in shape and energy to that seen in Figure 3a. This PIA likely corresponds to a transition between $1^1B_u^+$ and a state lying at ~3-3.5 eV, of $nA_g$ or CT character (Figure 5a,b). For a $1^1B_u^+$ symmetry $S_1$, the amount of $^1$(TT) character is expected to be greatly diminished. Performing pump-push-probe measurements as shown in Figure 5c on 'twisted' PDA with a push pulse centred at 940 nm ($t_{push}$ = 400 fs), however, results only in a depletion of the excited state population and no enhancement, with the positive $\Delta(\Delta T/T)$ signal rapidly decaying (~3.6 ps; Figure 5d). This timescale likely reflects a combination of electronic relaxation back to $1^1B_u^+$ and dissipation of excess energy on that surface consistent with timescales of thermalisation in PDAs[67]. This is strikingly different from the response observed in 'blue' chains, despite the chemical similarity (*e.g.* similar degree of sample alignment, Urbach disorder, *etc*; see SI, S3) between the two systems and the comparable total photon energy injected in the pump-push experiment. It further lends weight to the suggestion that the push pulse does not simply activate defect, charge, polaron and other branching pathways in 'blue' PDA. The observation that pushing a still singlet, $1^1B_u^+$ state does not result in population/lifetime enhancement, underlines the concept that it is the $^1$(TT) character within the $2^1A_g^-$ state of blue PDA that is the crucial factor, and that the appearance of the enhanced triplet PIA and lifetimes following the push result from separated triplet pairs.

The effect of the push pulse can thus be summarised as projecting the $2^1A_g^-$ wavefunction into a manifold of spatially separated triplet-pairs by using an optical perturbation that selectively couples the triplet-pair amplitudes of the $2^1A_g^-$ wavefunction to the $^1$(TT)* excited states, and their decay pathways towards nearly-free entangled triplet-pairs. Not only does this show how triplet-pairs can be harvested with low energy photons from non-luminescent polymers, it also provides a type of analysis of the total triplet character in the many-body $2^1A_g^-$ state, as well as the possibility to study real-time, real-space triplet motion in an organic material. In fact, as ultrafast time-resolved microscopy is rapidly emerging as a viable experimental techniques[95], it may even become possible in the near future to observe and manipulate the individual triplets in the nearly-free pairs.

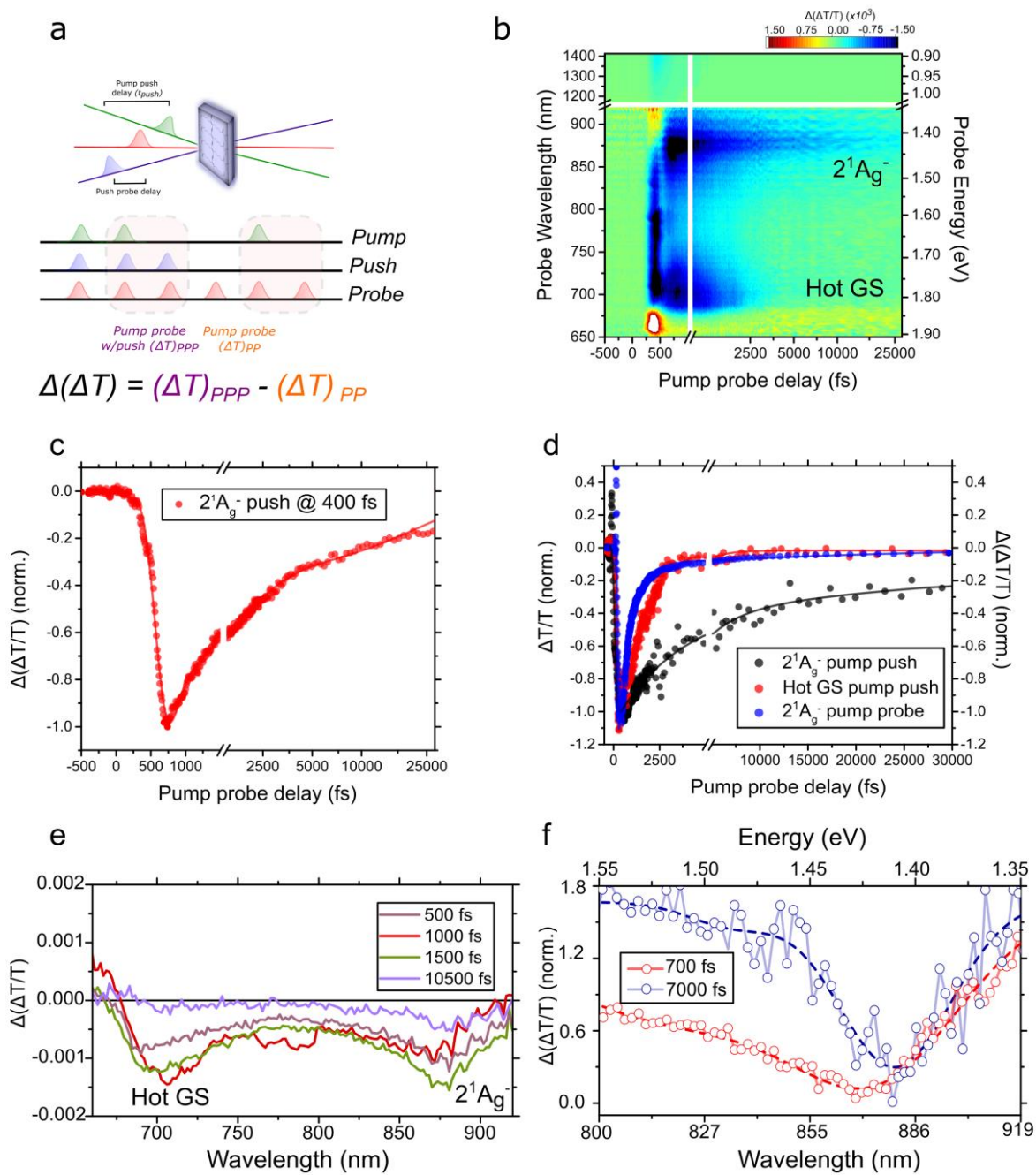

**Figure 4: Pump-push-probe spectroscopy of 'blue' polydiacetylene**: **a.** Pulse sequence used to record the electronic response of PDA chains with and without the push pulse (blue). The delay time of the push ($t_{push}$) after the pump pulse (green) and push pulse energy can be varied to examine their effects on the excited state dynamics (pump ~10 fs, push ~ 200 fs time resolution). **b.** Push-induced dynamics

of PDA with push pulse centred at 940 nm and $t_{push}$ of 400 fs. The negative signals indicate that the effect of the push is to increase the 'hot' ground state and $2^1A_g^-$ populations. **c.** Push-induced response of the $2^1A_g^-$ state when pushing at 400 fs. The $2^1A_g^-$ lifetime is increased to 15 ± 0.1 ps in this case. **d.** Comparison of decay (circles; solid line fit) for spectral region associated with the $2^1A_g^-$ state in pump-probe measurements, and $2^1A_g^-$ and 'hot' ground state in pump-push-probe experiments ($t_{push}$ = 200 fs; mapped back to single scale where all pulses arrive at $t$ = 0 fs). The PIA associated with $2^1A_g^-$ rapidly decays in pump-probe experiments; application of the push however enhances the lifetime of this PIA by at least one order of magnitude. **e.** Spectral cuts of pump-push-probe spectrum. The 'hot' ground state and $2^1A_g^-$ PIAs are present almost immediately after the push pulse with the latter being slightly delayed. The data is shown for $t_{push}$ = 400 fs, with the times in the legends indicating the pump-probe delay. **f.** Magnification of the spectral region associated with the pushed $2^1A_g^-$ PIA; the spectra are normalised at 900 nm. Dashed overlay is a fit of two Gaussians to the profile (see SI, Figure S10 for further details) and highlights the narrowing of the spectrum at longer time delays following the push.

Having ruled out other potential origins of the push-induced signals we can return to consider the mechanism of triplet pair extraction from $2^1A_g^-$. The $2^1A_g^-$ manifold is energetically broad, so we examined the effect of varying the push energy (SI, Figure S10) at a constant $t_{push}$ of 400 fs. When tuning our push pulse to be resonant with either the high- or low-energy side of the $2^1A_g^-$ PIA at 860 nm and 1100 nm respectively we observe no change in the dynamics as compared to a 940 nm push, with the same push-induced population change and aforementioned lifetime enhancement. However, when we centre our push pulse at 1350 nm in the tail of the $2^1A_g^-$ state no push-induced signal is observed. Based on the absorption cross-section at 1350 nm, obtained from the $\Delta T/T$ signal in Figure 3, and relative push power used at 1350 nm compared to 940 nm we would expect a maximum $\Delta(\Delta T/T)$ of ~4 × 10$^{-4}$, well above our noise floor of ~2 × 10$^{-5}$. This suggests that the 1350 nm push does not extract separated triplets from the $2^1A_g^-$ superposition. We note that the total energy provided *via* the 1350 nm push is ~2.32 eV ($2^1A_g^-$ at ~1.4 eV [64] + 0.92 eV from push), which is significantly higher than the expected energy for 2 triplet (T) excitons at ~1.8 – 2.0 eV [64]. Given that the energy difference between the 1100 nm and 1350 nm push pulses is only ~200 meV, it suggests that in both instances a 'hot' $2^1A_g^-$ state will be formed, but only for the (860 -) 1100 nm push is there sufficient energy to access $^1(TT)^*$ states. Hence, contrary to the original model proposed in the literature[39], a 'hot' $2^1A_g^-$ state is not a sufficient condition to extract spatially separated triplets from the mixed $2^1A_g^-$ wavefunction[49,68], and in addition projection of the wavefunction *via* excitation into the triplet-pair manifold is required. More interestingly given that regardless of the push energy (and time delay) only the lifetime of $^1(TT)$ contribution can be greatly enhanced, the results indicate that not all components of the $2^1A_g^-$ superposition (*e.g.* CT or localized exciton) can likely be addressed. There are likely to be several contributing factors. These other amplitudes (Figure 1b) constitute a relatively small fraction of the $2^1A_g^-$ wavefunction and consequently would contribute only weakly to the total excited-state absorption[56]. Indeed, the similarity of the PIA lineshape to that expected of triplet-pairs also suggests the triplet-like character dominates the transition[56]. Moreover, unlike the transition $^1(TT)^*$-$^1[T…T]$, the one-electron components do not have an appropriately placed manifold of states to cross into and/or net entropic gain on transition, hence would only contribute to the sub-400 fs depletion signals observed in the pump-push-probe spectra (Figure 4d).

We further highlight the unexpected observation that the push photon energy does not appear to alter the degree of $^1$[T…T] separation (SI, Figure S10). This result indicates that formation of near-free triplet-pairs occurs from the lowest energy $^1$(TT)* state. Any excess energy in the push pulse beyond the $^1$(TT)* threshold is rapidly (sub-200 fs) lost as heat/vibration as the 'hot' $^1$(TT)* relaxes, making the subsequent dynamics independent of the initial photon energy. This makes a surprising contrast with conventional acene thin-film systems, where $^1$(TT) separation is known to be thermally activated[96], or other conjugated-polymer materials where excess photon energy enhances the triplet yield[18,45,51,97], and it points to a crucial difference in the nature of our experiment. The previously reported photon energy dependences explore the importance of excess vibrational energy in the initial bright state as it branches between $2^1A_g^-$ versus $^1$[T…T], and the excess 'heat' deposited in $2^1A_g^-$ under such conditions was not found to result in further $^1$[T…T] formation. Similar effects are reported in some PDAs as well[68], but the absence of a 1-photon excitation energy dependence in the present derivative (Figure 2c) suggests this pathway is not important here. We instead deposit excess energy directly in $2^1A_g^-$, a well-defined electronic state in polyenes with a binding energy significantly above thermal energies. If, as our experimental results suggest, the push pulse projects the system into an area of the potential energy surface where relaxation is branched between $2^1A_g^-$ and $^1$[T...T], there is *a priori* no reason why that branching should involve any kind of thermal barrier[79]. Indeed in no case do we observe a clear rise of the $^1$[T...T] signature (distinct from the $2^1A_g^-$ decay profile, as discussed above), nor are there any identifiable signatures of $^1$(TT)*. Consequently, if $^1$(TT)* has no appreciable lifetime, there is no reason for vibrational relaxation within this state to have any effect on the dynamics. Based on this preliminary scan of push energy, the crucial parameter that determines the likelihood of $^1$[T…T] is thus not the degree of vibrational excitation or 'heat' – as found to be the case in many intermolecular singlet fission systems[14,82,98,99] – but the underlying nature of the upper $^1$(TT)* state. We suggest that this highly electronically excited state is more delocalized than $2^1A_g^-$, a common phenomenon in conjugated polymers[100]. Coulombic repulsion between a more dipolar T* exciton, created by the push, and more tightly bound electron/holes of the T exciton, results in greater average T-T separation enabling relaxation into the overall more localized $^1$[T…T] state.

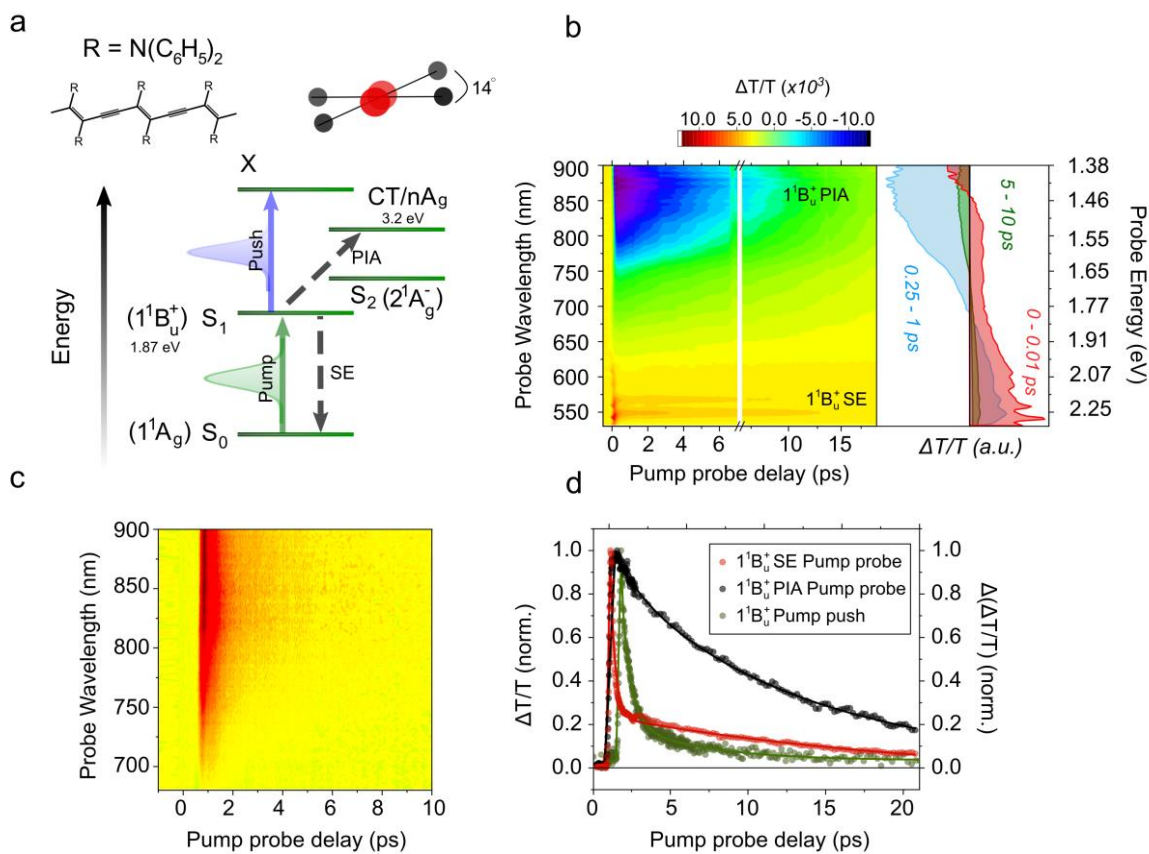

**Figure 5: Pump-probe and pump-push-probe spectroscopy of 'twisted' polydiacetylene: a.** Chemical structure (top) and energy diagram of 'twisted' PDA. Twisting the polymer backbone (14° torsional angle) pulls the $2^1A_g^-$ state above $^1B_u^+$, resulting in a bight $S_1$. Pump and push pulse are shown in green and blue, respectively. Dashed arrows indicate transitions observed in the pump-probe spectra, from which an estimate of the energy of various states is garnered. **b.** Pump-probe spectrum of 'twisted' PDA. Sharp SE peaks are observed between 520-600 nm, with a broad $1^1B_u^+$ PIA between 750-900 nm. **c.** Pump-push-probe spectrum of 'twisted' PDA. Pushing on the $1^1B_u^+$ PIA ($t_{push}$ = 400 fs) results only in a positive depletion signal. **d.** Comparison of pump-probe and pump-push probe kinetics (circles raw data; solid lines fit). The SE and PIA show multi-exponential decays with $\tau_{av}$ ~2.1 ps and $\tau_{av}$ ~13.8 ps, respectively. The push depletion signal (green) decays over ~3.6 ps.

In summary, the precise mechanism for transition between the $2^1A_g^-$ and $^1[T…T]$ manifolds is challenging to ascertain. This is because of the short-lived nature of the intermediates, which cannot be spectrally resolved or separated by external stimuli such as a magnetic field, even at low temperatures (SI, S4). The lack of push time and energy dependence indicates that following the push, population is driven from $2^1A_g^-$ into $^1(TT)^*$. There is a configurational barrier associated with reaching this more T-like surface, as demonstrated by the fact that push energies below ~1.1 eV do not generate $^1[T…T]$ like signatures. The instantaneous rise of the 'hot' GS signal following the push, suggests energy is rapidly dissipated from $^1(TT)^*$ as population cools to $2^1A_g^-$. Coulombic repulsion between the more delocalised, dipolar T* electron-hole pair and T, where the electron and hole are more tightly bound, is likely a large driving force for rapid separation. The T* excitons will effectively scatter with phonons due to the strong exciton-phonon coupling in PDA, resulting in their rapid relaxation (within our time resolution)[101]. The large phase space that the $^1$(TT)* state can explore will also entropically drive triplet-pair separation. Given we observe no long-lived triplets and geminate triplet-triplet annihilation remains rapid it is likely no other spin states *e.g.* $^3$(TT) or $^5$(TT) are involved. This mechanism is potentially matched by recent high-level calculations on polyenes[102,103]. These calculations identify a higher-lying $B_u^-$ state with similar spin entanglement to $2^1A_g^-$, but no binding energy and less mixing with other states. This state could effectively be $^1$[T…T], suggesting the upper surface ($B_u$ character) has some parallel relaxation to $2^1A_g^-$ and $B_u^-$, which are electronically and spin-wise similar and located at comparable energies. However, further material specific calculations are necessary to understand and corroborate this mechanism. In general calculations on a highly correlated many-body state such as $2^1A_g^-$ and $^1$[T…T] are towards the limit of current theoretical methods (*e.g.* Quantum Monte Carlo), requiring the simulation of an effective model, rather than a truly *ab initio* approach. Moreover, in a polymeric system of the size of PDA, even model-based computations rapidly become intractable due to the exponentially growing number of electronic configurations. Consequently, future work should be targeted at developing new theoretical methodologies for tackling such problems.

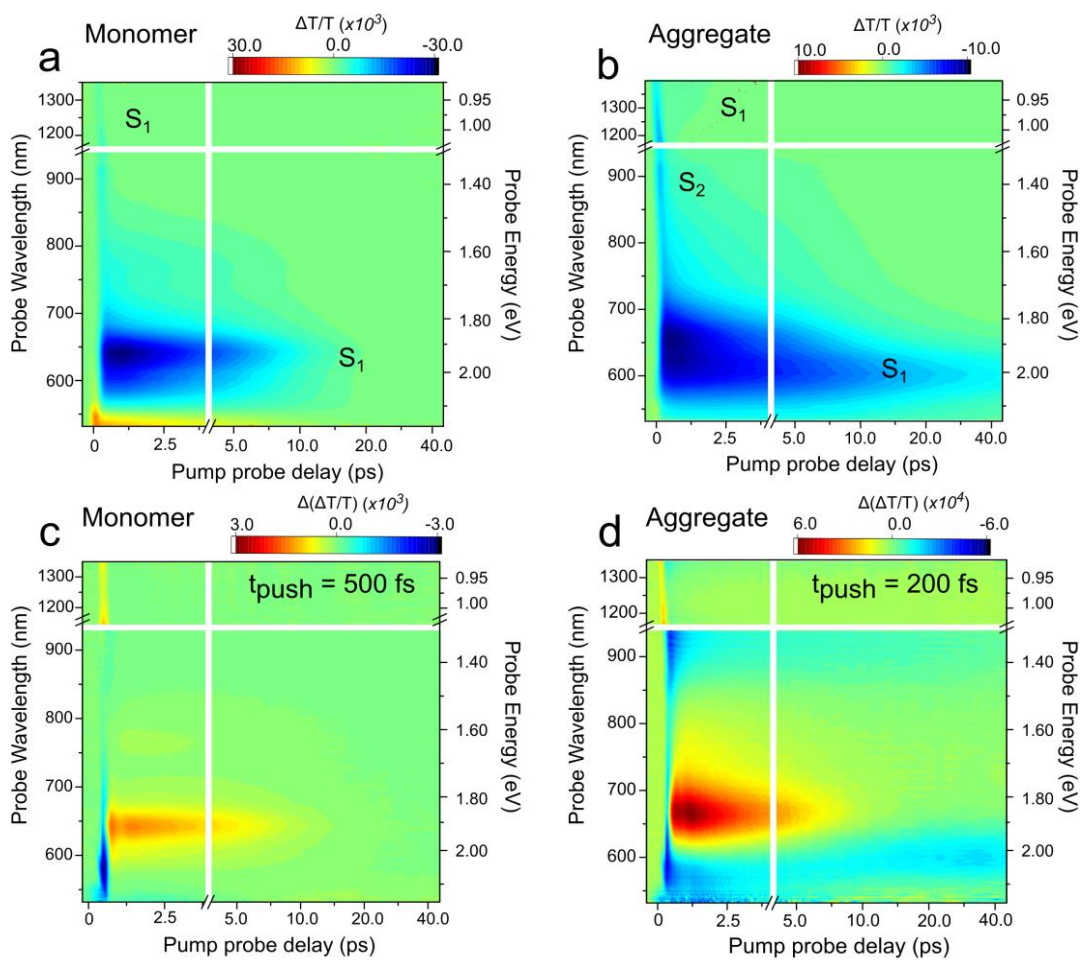

**Figure 6: Pump-probe and pump-push-probe spectroscopy of a carotenoid (astaxanathin, AXT) monomer and weakly coupled H-aggregate. a.** Differential transmission map following excitation of isolated AXT at 515 nm (~200 fs time resolution). The PIA bands centred at 620 nm and 1220 nm are assigned to the $S_1$ state, which in the AXT monomer is $2^1A_g^-$ in character. **b.** Pump-probe spectrum of H-aggregated AXT. The PIA bands at 620 and 1220 nm correspond to $S_1$, which on the timescales considered (0.3 – 10 ps) have a characteristic of the strongly bound $2^1A_g^-$ states seen in the monomer. In both the monomer and aggregate the 1220 nm $S_1$ PIA is ~10 times weaker and strongly overlapped at early times (sub-200 fs) with $S_2$, but normalisation of the kinetics shows it precisely follows the same decay as the PIA at 620 nm (SI, Figure S13). **c.** Pump-push-probe spectrum of isolated AXT following a push pulse at 1220 nm and $t_{push}$ = 500 fs where the $S_1$ population is at a maximum. Only depletion of $S_1$ is observed (positive $\Delta(\Delta T/T)$ response 600 – 700 nm). **d.** Push-induced response of AXT aggregates following 1220 nm push pulse at 200 fs. Initially there is a depletion of the population at 620 nm; this decays into a negative 'enhancement' that is still increasing at the end of the measurement window (40 ps).

**Carotenoids** Carotenoids are arguably some of the most important naturally occurring photoactive organic molecules. Commonly found bound to the proteins of photosynthetic organisms they play a key role in light-harvesting and photoprotection in plants and bacteria, as well as giving rise to the characteristic colours of many crustaceans and fish[104]. The electronic structure of carotenoids (and their aggregates) is remarkably similar to that of 'blue' PDA as shown in Figure 1a. Additionally, weakly (both structurally and electronically) bound carotenoid aggregates have been shown to undergo efficient interchain singlet fission with triplet yields up to 200%[52,53], and the introduction of comparable polyene character is increasingly suggested as a design strategy for intramolecular singlet fission[18,26,45,55,105–107]. We use the carotenoid astaxanthin (AXT) to probe the analogy with the PDA $2^1A_g^-$ state, both in isolated small molecules and aggregates. These were prepared following reported protocols[52] and exhibit ground-state absorption edges at ~530 nm (monomer) and ~550 nm (aggregate) consistent with literature (SI, Figure S10).

In Figure 6a,b we present the pump-probe characterization of monomeric AXT and its weakly coupled H-aggregate, excited at 515 nm (pulse duration ~200 fs). Longer pulses with a peak-power of ~0.1 × $10^6$ W cm$^{-2}$ (10 times lower peak power than with PDA) were used to avoid sample degradation/isomerisation. In any case the 200 fs duration of our push pulse means that three-pulse experiment is not sensitive to any short-lived intermediate states that may be present. In agreement with previous reports on AXT[52] and similar carotenoids[108] we assign the prominent excited-state absorption band at 600 – 850 nm and the markedly weaker tail 1200 – 1350 nm in the monomer to the $S_1$ (*i.e.* $2^1A_g^-$) state, which is formed from the initial $S_2$ (SE at ~550 nm) in <300 fs. The aggregate presents similar, albeit slightly shifted features, where once again the initial $S_2$ state, which has a weak PIA in the 850 – 1100 nm region, is depleted within 300 fs. The long-lived PIA peak at ~600 nm has previously been assigned to triplet-pairs formed through singlet fission[52]. Here we highlight that on intermediate timescales (0.3-10 ps) the spectrum is broadened and red-shifted and exhibits substantial rapid decay (SI, Figures S11-S13). These features are characteristic of the strongly bound $2^1A_g^-$ states observed in monomeric AXT and 'blue' PDA, and we thus propose the state on these timescales is best represented by a superposition as in Figure 1b with primarily triplet-pair character and similarly denote it as $S_1$. In both aggregated and monomeric AXT the $S_1$ PIA also has contributions in the near- infrared (1200 – 1350 nm)[52]. Here the PIA is ~10 times weaker than in the visible and at early times overlapped with the strong and rapidly decaying $S_2$ PIA; normalizing the kinetic at 620 nm and 1220 nm however shows the two bands decay at the same rate (SI, Figure S13). Depending on the number of conjugated double bonds several other states have been shown to sit in the $S_2$-$S_1$ gap of carotenoids[109]. In AXT for instance, another triplet-pair state $1^1B_u^-$ is predicted to lie below the $1^1B_u^+$ ($S_2$) transition, albeit with more covalent character and lower oscillator strength especially in the solution phase[40,110]. The role of this and other intermediate states has been intensely debated[111–113], however given the ~200 fs pulse duration, we presume rapid decay from these states to have occurred within our resolution and before arrival of the push pulse ($t_{push}$ > 200 fs in both monomeric and aggregated AXT, where models of dark-state-mediated internal conversion in carotenoids agree that the bulk population resides in $2^1A_g^{-52}$). We note that previous measurements with faster, sub-30 fs pulses, were also unable to observe these or any other dark states, concluding they did not influence the electronic dynamics of AXT[52]. Consequently, we base our observations on a three state model considering only $1^1A_g^-$ ($S_0$), $2^1A_g^-$ ($S_1$) and $1^1B_u^+$ ($S_2$) [111,114–116] (further discussion is provided in SI, Figure S12).

To probe the superpositions contained within the $S_1$ state, pump-push-probe experiments were performed on both samples with a push pulse centred at 620 nm and 1220 nm; the $t_{push}$ was chosen to overlap with the maximum in the rise of the respective PIAs (the behavior is approximately independent of the precise $t_{push}$ values, further emphasizing the role of any rapidly decaying intermediate states to be minimal, SI, Figure S14, or the choice of push wavelength within the indicated bands, SI, Figure S15). Figure 6c shows the push-induced signal following a push at 1220 nm of isolated AXT. Pushing the $S_1$ ($2^1A_g^-$) PIA band results in a positive push-induced $\Delta(\Delta T/T)$ signal indicating population has been driven out of $S_1$ to higher lying states, with the depletion signal decaying over ~5 ± 0.05 ps, in accord the with the intrinsic $S_1$ lifetime. In this case no (long-lived) enhancement can be observed which suggests that on a single chain there is insufficient space for triplet-pairs to separate from the $2^1A_g^-$ superposition. In the case of the aggregates the behavior is markedly different as shown in Figure 6d. Immediately following the push pulse there is a depletion of $S_1$, resulting in a positive $\Delta(\Delta T/T)$ signal that decays ($\tau$~4.2 ± 0.05 ps) to yield a slightly blue shifted negative $\Delta(\Delta T/T)$ 'enhancement' signal, that grows to the end of the measurement window (SI, Figure S13). The instrument resolution precludes measurement to longer pump-probe time delays, but previous studies show no significant deviation from the $^1(TT)/2^1A_g^-$ manifold until the onset of bimolecular triplet-triplet annihilation beyond 1 ns[52]. Pump-push-probe measurements to this timescale would likely provide insight into the triplet transport and recombination dynamics but this is reserved for future study. As with PDA, this growth in the negative $\Delta(\Delta T/T)$ signal indicates that the push slows the decay of the $S_1$ state, *i.e.* the population is able to reach a new, more stable part of the configuration space. We note at the pump and push fluences used here, the transition between $2^1A_g^-$ and the higher excited state follows a strict one-photon dependence in both monomers and aggregates (SI, Figures S16-S17). This is in contrast to other studies which have shown access to the triplet state manifold *via* 3-4 photon excitation from the $1^1A_g^-$ ground state of monomeric carotenoids[92,93].

In other monomeric carotenoids long-lived signatures of population *depletion*, similar to those seen in Figure 6c, have been observed following application of a push-pulse to the $2^1A_g^-$ PIA[117,118]. However, to the best of our knowledge, in none of these systems has there been evidence for an enhancement in the lifetime of the triplet-pair like (or other) contributions or branching into a particular dark-state. In *monomeric* β-carotene it has been reported that excitation from the ground state with significant excess photon energy can result in a longer-lived PIA band, blue-shifted from the $S_1$-$S_n$ transition[118], and the same effect is evident in several other carotenoids[119]. This bears superficial resemblance to our push-induced effects in that an enhancement signal is also observed in 530 – 570 nm range in Figure 6d, but caution should be taken in comparing spectra between different carotenoids. Recent work by Ostroumov *et al.* has shown that long-lived features in the transient absorption spectra of carotenoids such as β-carotene can be explained by population impurities arising from cis-isomers in the sample[114]. Following rigorous purification, these long-lived features are removed and the entire transient absorption spectrum after 1 ps can be explained by single $S_1$-$S_n$ transition[114,115]. Hence, care must be taken in interpreting results on these materials, especially given such *cis*-isomers can be generated transiently at high-excitation densities and pump energies[120]. It is for this reason we specifically chose to excite towards the lower energy band-edge of AXT at 515 nm (2.4 eV) as opposed to at 400 nm (3.1 eV) as is more commonly used in the literature. In addition to these precautions, our control measurements on monomeric AXT (Figure 6a,c) reveal no long-lived features demonstrate that the push pulse does not access any other electronic pathways. Likewise, control measurements on monomeric β-carotene (SI, Figure S17), where previous studies suggest intervening 'dark' states may have a more significant role on the photodynamics[92,121,122], reveal only depletion of the $S_1/2^1A_g^-$ state with no longer-lived features. Pathways involving multiple dark states, as reported by Larsen *et al*. applying a similar total excitation energy to monomeric β-carotene[118] in the ground electronic state, do not appear to be active in our systems following selective stimulation of the $S_1$-$S_n$ transition.

Introducing an increasing number of conjugated oxygen atoms into the carotenoid backbone has also been shown to make these molecules more robust to conformational effects[104], which guided our choice of AXT as a model carotenoid. Performing absorption measurements after each pump-push-probe scan further confirms there are no structural changes in the molecule during measurements (SI, Figure S12). The high-energy edge of the $S_1$-$S_n$ PIA (550 – 570 nm feature in Figure 6b) has been assigned to both a dark S* state and *cis*-isomer impurities in some monomeric carotenoids and protein complexes with isolated carotenoids[118,119,123,124], but these have not been found to contribute significantly to the spectrum of AXT. Nonetheless, it is possible that the early-time (<1 ps) negative Δ(*ΔT/T*) feature we detect in Figure 6d reflects a slight enhancement of the lifetime or population of such dark/impurity states. However, this feature exhibits rapid decay which is not matched by corresponding growth in the band at 600 nm assigned to the $2^1A_g^-/^1(TT)$ state in carotenoid aggregates[52,79] (SI, Figure S13). Any push-induced effects in possible dark/impurity states thus have no connection to the 'freeing' of triplet-pairs. In light of this, the conformational robustness of AXT, the control experiments detailed in the supporting information (SI, Figure S12 – S15) and the similarity of push-induced effects between AXT aggregates and PDA, where we expect no such isomerization effects, we are confident that our results are not related to population impurities.

In the study of monomeric carotenoids, comparable PIA features are also frequently analysed in terms of other dark states such as $1^1B_u^-$, which are also predicted to lie below the bright $1^1B_u^+$ state for sufficiently long molecules[39,40]. Such states are invoked both to mediate internal conversion into $2^1A_g^-$ and to explain long-lived features in the transient spectra, though their assignment remains highly controversial[109,118], and the demonstration of a conical intersection between $1^1B_u^+$ and $2^1A_g^-$ suggests no intervening states are present[111]. In the present work, the limited temporal resolution of the push experiment means that only the terminal $2^1A_g^-$ is detectable, and we cannot comment directly on the presence of other dark states within the relaxation pathway from high-lying $S_n$. Within the aggregates, it has previously been suggested that $1^1B_u^-$ might play a role in the SF process[52], as it is also expected to carry triplet-pair character[39,40]. It is certainly possible that the $^1[T…T]$ we observe is related to $1^1B_u^-$, though it is beyond the scope of this work to determine the exact symmetry of the state. Moreover, we note that these symmetry labels are strictly defined only for polyacetylene. The electronic coupling within multi-chromophoric aggregates must alter the symmetry of the system, such that it is unlikely that precisely the same states can be used to describe monomer and aggregate electronic structure. This being the case, we consider it most appropriate to focus in this work on the relevant diabatic contributions to the overall states, *e.g.* $^1(TT)$ and $^1[T…T]$.

Nonetheless as for PDA, it is important to briefly consider other origins of the observed signals in Figure 6c,d. The use of band-edge excitation, with a pump pulse of relatively low peak power, allows the main 'defect' contribution of *cis*-isomers to be ruled out[119]. Signatures of intramolecular charge-transfer (ICT) states have been observed in some carotenoid monomers, including AXT[125] and in other carotenoids *e.g.* peridinin, it has even been concluded that $S_1$ has a degree of charge-transfer character. Pushing the $S_1$ state of peridinin in polar solvents activates these ICT states driving population into the $nA_g$ manifold[126]. However, in all cases the lifetime of these ICT states has been shown to be ~5-10 ps, much shorter than the push induced lifetime enhancement observed in Figure 6d. In the case of AXT (and other carotenoid aggregates), ICT features have yet to be investigated. In order to test for CT states in AXT aggregates, pump-push-probe measurements were repeated, with aggregates dissolved in a less polar solvent, acetone (Figure SI, S15). As the solvent polarity decreases, it is expected that CT states will be relatively destabilised with a shorter rise and decay time. Although we observe a slight blue-shift of the spectral features in pump-push-probe measurement in acetone with respect to measurements performed in DMSO (in line with the expected solvatochromism of AXT[52]), no significant changes are observed in the push induced kinetics. This suggests that the effect of the push in AXT aggregates is not to drive population into a long-lived charge transfer state. We also note that based on spectroelectrochemistry measurements of AXT any PIA features associated with charged states would be expected to sit in the near-infrared, as opposed to ~600 nm[127].

Having ruled out, ICT, defect, dark, *etc* states, the remarkable similarity in behavior to PDA suggests that in AXT aggregates, optical stimulation of $S_1$ (which has some $2^1A_g^-$ superposition state contribution on early timescales) separates and delocalizes the bound triplet-pairs over multiple chains to a greater degree than occurs naturally within the aggregates. The slow growth of the depletion signal and additional push induced features (*e.g.* depletion signal at 800-900 nm in Figure 6d) suggests that the amount of $^1[T…T]$ optically liberated from $2^1A_g^-$ in AXT aggregates is less. This is in line with the weak interchromophore coupling and potential additional branching pathways in carotenoids[59,109,128]; this also makes reliable kinetic modelling of such spectra challenging. Although AXT is chemically different to PDA, the likeness in photophysical response suggests that the possibility for extraction of the $^1[T…T]$ state from a $2^1A_g^-$ superposition, *via* an optical pulse, is widespread. The range of energetic orderings, numerous 'dark' states and potential for conformational changes on photoexcitation means further investigation is required to show these observations hold for all carotenoid systems.

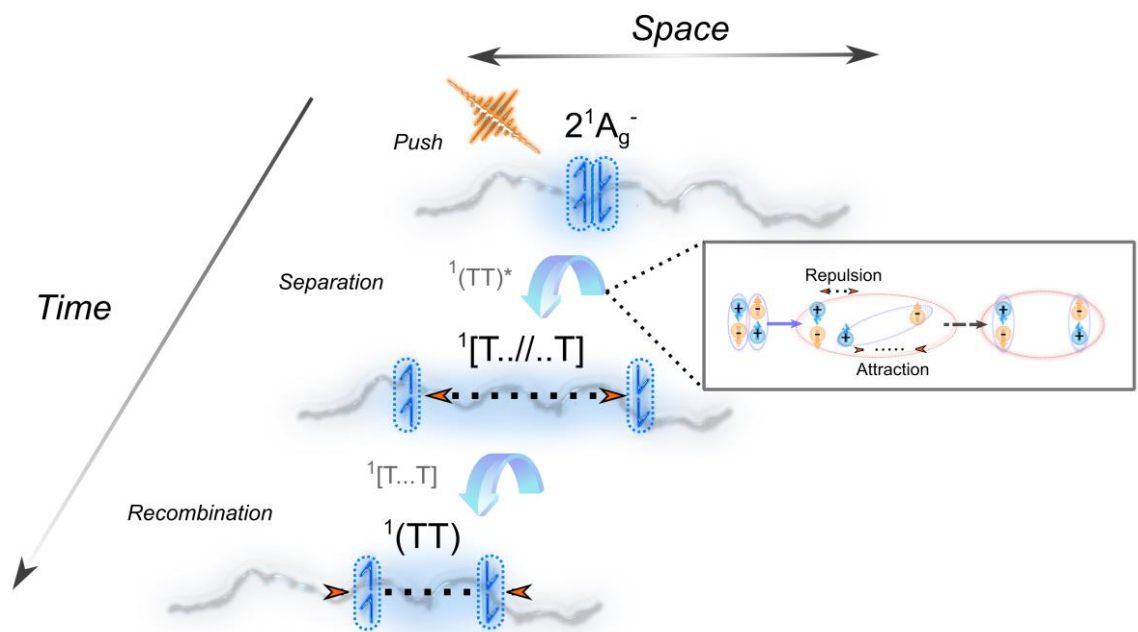

**Figure 7: Schematic of the spatial separation of triplet-pairs from the $2^1A_g^-$ state in pi-conjugated molecular systems.** The bound triplet-pair excitons (dashed boxes on blue background) separate along the conjugated molecular backbone into a spatially free triplet-pair state where there is maximum *spin* entanglement but minimum *spatial* entanglement. After a given time, the spins recombine to give rise to the observation of an overall lifetime enhancement in $2^1A_g^-$. The inset (black box) shows a suggested mechanism for triplet-pair separation. The push pulse acts on one of the triplets in $^1(TT)$ to generate an excited T* state where the electron and hole are spatially separated. This results in strong coulombic repulsion between the dipolar T* exciton and electrons/holes in T, which drives spatial separation within the triplet-pair state.

Conclusion

We have provided experimental evidence that the $2^1A_g^-$ state in polydiacetylene and a carotenoid aggregate can be a superposition state with significant $^1(TT)$ character. More importantly we have demonstrated that optical excitation addressing the triplet-pair amplitudes of the $2^1A_g^-$ wavefunction, can be used to project the $2^1A_g^-$ wavefunction into a manifold of spatially separated triplet-pair states. The resulting lifetime enhancements observed in the $2^1A_g^-$ excited state absorption (PIA) is due to long-range spatial distribution of the spin-entangled triplet-pairs along a PDA polymer chain or between carotenoid molecules in weakly coupled aggregates (Figure 7). The degree of electronic coupling between the individual monomer chromophores dictates the time taken for spatial separation of these pairs and the overall yield liberated from $2^1A_g^-$. In the strongly coupled, highly ordered PDA, stimulated separation into $^1[T…T]$ takes ~200 fs, whereas in the weakly bound carotenoid aggregates the spatial separation happens over ~5 ps and the yield of separation is suggested to be smaller. In small monomers of the carotenoid AXT there is no evidence for extraction of $^1[T…T]$ by the one-photon push pulse due to the strong spatial confinement. By exploring complementary polyene systems (*e.g.* 'twisted' PDA, β- carotene monomers, *etc*) and through a plethora of controls (*e.g.* solvent, temperature and defect density dependence) we are able to robustly rule out many other signatures which often complicate the interpretation of excited state dynamics in polyenes. Push time and energy dependence measurements additionally allow the determination of a mechanism for $2^1A_g^-$ to $^1[T…T]$ conversion, whereby following the push pulse there is rapid vibrational relaxation to a delocalised $^1(TT)^*$. Here, strong coulombic repulsion drives triplet-pair separation; this mechanism has parallels with recent high-level theoretical calculations on polyenes[103].

Our results indicate that it is possible not only to 'control' but enhance and quantify the $^1(TT)$ character of the $2^1A_g^-$ wavefunction providing an all-optical approach to steering the singlet-fission reaction towards the desired triplet photoproduct over internal conversion. Furthermore, the indirect evidence present for the near-free triplet pairs remaining entangled over their entire lifetime ($S_1$ lifetime enhancement decays only *via* recombination of geminate pairs to the $S_0$) will – we hope – inspire future experiments that tackle the challenging task of measuring such quantum correlations in organic materials[32].

Given the prevalence of $2^1A_g^-$ states in molecular systems we expect our results to be applicable to a wide range of other technologically relevant systems such as polyacenes, iso-indingo-based polymers and other singlet fission materials. Further studies should aim to address the universality of our observations potentially using techniques such as 2D electronic spectroscopy which may be sensitive to 'dark state' signatures in carotenoid systems[59,128]. In fact, it may be possible to develop a quantitative tool based on the general idea of optically projecting out certain contributions to a complex wavefunction, For example, by quantifying the degree of $^1$(TT) character in materials possessing a S$_1$ ($2^1A_g^-$) state, it could become possible to rationally design new polymers for singlet fission and photoprotection by identifying structures that optimise the $^1$(TT) contribution to $2^1A_g^-$. Indeed, it has been shown extensively that control over the coupling and energy of CT-like excitations can influence singlet fission rates and the energy of $^1$(TT) states. Mixing CT character *via* donor-acceptor chemistry can lead to intramolecular singlet fission and one often needs to avoid the formation of 'dark' $2^1A_g^-$ states to achieve good fission yields[10,45,61,81,129–131]. In contrast our results show that by manipulating the $2^1A_g^-$ state efficient triplet production can in-fact be achieved. We control the $^1$[T...T] lifetime externally, not by materials choice – although the two could be combined – and produce triplet pairs that can be created and separated on demand; an altogether different approach for achieving efficient intramolecular fission. More generally, given the ubiquity of 'hybrid' states in organic materials, particularly those mixing charge transfer and excitonic characters, mean this technique of projecting out the individual components will provide a powerful tool to explore and manipulate electronic states in many other molecular systems.

Experimental Methods

Sample preparation

'Blue' Polydiacetylene

The 3BCMU (3-methyl-n-butoxy-carbonylmethyl-urethane) diacetylene molecules were synthetized in-house using the method previously outlined by Se *et al.* and references therein[132]. Single crystals were grown by slow crystallization in a diacetylene-methyl iso-butyl ketone supersaturated solution at 4°C in the dark. Platelets with an area of a few square millimetres and a thickness around 100-200 μm are typically produced. To avoid photo- and thermal- polymerisation and to keep the polymer content as low as possible, samples were stored in the dark below -20° prior to measurement. Typical polymer contents by weight are then in the $10^{-4}$ - $10^{-3}$ range. The intrachain separation (~100 nm; homogeneous distribution) is obtained from the absorption optical density ($OD$) and Beer Law, where $OD = \frac{\propto \times l \times c}{2.3}$ where $\propto$ is taken as $\sim 1 \times 10^6$ cm$^{-1}$ for light polarized parallel to the long axis of the chains and $l$ is ~150 μm. The spatial separation between chains is assumed to be significantly large enough that there are negligible inter-chain interactions.

'Twisted' Polydiacetylene

Crystalline monodomains of the 'twisited' diacetylene monomer were first grown before being slightly polymerized. Large size crystalline regions (1-2 mm$^2$) were obtained using a melt processing method. A few mg of the purified diacetylene powder were melted (T ≥ 80 °C) and the liquid injected, using capillarity action, in the space between two superimposed microscope coverslips. After rapid cooling the thin liquid film crystallization led to the formation of a highly polycrystalline structure, being the assembly of microdomains. The sample was then heated again under a polarized optical microscope until almost all the melting had taken place with the preservation of a few single crystal germs. At that point the sample was cooled again at a very slow cooling rate to induce the growth of hundreds of μm – mm scale domains from the germs until room temperature was reached. Polymer chains were then generated by exposure to X-ray light in a diffractometer (Rigaku, Smartlab)[94,133]. The polymer content was adjusted by trial and error and kept low enough (typically below 0.1 % in weight) to form a solid solution of isolated 'twisted' polydiacetylene chains inside their diacetylene host crystal.

Astaxanthin

Racemic astaxanthin (AXT) was generously donated by BASF. AXT monomer solutions were prepared at a concentration of 100 μM in DMSO and heated at 50 °C until clear. The AXT aggregate was prepared by mixing a 1000 μM solution of AXT in DMSO at 80 °C in a 1:9 ratio with water at 80 °C[52]. Samples were measured in a 1 mm pathlength cuvette (Hellma) with an optical density of 0.3.

Absorption spectroscopy

Polarised absorption spectroscopy of PDA crystals was performed using a home built setup with a white light source generated by focussing the fundamental of a Yb-based amplified system (PHAROS, Light Conversion) into a 4 mm YAG crystal. The resulting absorption (corrected for the sample substrate) was then collected by imaging with a Silicon photodiode array camera (Entwicklunsbüro Stresing; visible monochromator 550 nm blazed grating). The polarisation of the white light was controlled by means of a half waveplate (Eksma). The absorption spectrum of AXT was performed using a commercial PerkinElmer lambda 750 UV–vis–NIR setup. A Xe lamp was used as the light source, and all measurements were performed under standard ambient conditions. The spectra were measured simultaneously with the solvent to correct for its absorbance in case of the carotenoids.

X-Ray diffraction

X-Ray diffraction (XRD) was performed using a Bruker X-Ray D8 Advance diffractometer with Cu Kα1,2 radiation ($\lambda$ = 1.541 Å). To prevent sample damage from X-Rays, measurements were carried out at 12 K using an Oxford Cyrosytem PheniX stage. Spectra were collected with an angular range of $1 < 2\theta < 56°$ and $\theta = 0.0051°$ over 60 minutes. Measurements were made on a flat crystal mounted in a non-specific orientation. The Bruker Topas software was used to carry out Le Bail analysis over an angular range of $8.75 < 2\theta < 26.7°$. Backgrounds were fit with a Chebyshev polynomial function and the peak shape modelled with a pseudo-Voigt function.

Femtosecond pump-probe and pump-push-probe spectroscopy

The fs-TA experiments were performed using a Yb-based amplified system (PHAROS, Light Conversion) providing 14.5 W at 1030 nm and 38 kHz repetition rate. The probe beam was generated by focusing a portion of the fundamental in a 4 mm YAG substrate and spanned from 520 nm to 1400 nm. The pump pulses were generated in home-built noncollinear optical parametric amplifiers (NOPAs), as previously outlined by Liebel *et al.*[134]. The NOPAs output (~4 to 5 mW) was centred typically between 520 and 560 nm (FWHM ~65-80 nm) depending on the exact experiment, and pulses were compressed using a chirped mirror and wedge prism (Layerterc) combination to a temporal duration of ~9 fs. Compression was determined by second-harmonic generation frequency-resolved optical gating (SHG-FROG; upper limit) and further confirmed by reference measurements on acetonitrile where the 2200 cm$^{-1}$ mode could be resolved (see SI, Figure S7). The probe white light was delayed using a computer-controlled piezoelectric translation stage (Physik Instrumente), and a sequence of probe pulses with and without pump was generated using a chopper wheel (Thorlabs) on the pump beam. The pump irradiance was set to a maximum of 30 μJ/cm$^2$. After the sample, the probe pulse was split with a 950 nm dichroic mirror (Thorlabs). The visible part (520–950 nm) was then imaged with a Silicon photodiode array camera (Entwicklunsbüro Stresing; visible monochromator 550 nm blazed grating). The near infrared part was imaged using an InGaAs photodiode array camera (Sensors Unlimited; 1200 nm blazed grating). Measurements were carried out with a time step size of 4 fs out to 2 ps to minimize the exposure time of the sample to the beam. Unless otherwise stated, all measurements were carried out with the probe polarisation set parallel with respect to that of the pump (using a half-waveplate; Eksma) and along the PDA chains. The absorption spectrum of samples was measured after each pump-probe sweep to account for any sample degradation.

For the pump-push-probe spectroscopy, an additional third pulse was added to the above configuration, and spatially and temporally overlapped in the sample through a boxcar geometry. The pump-push delay of the push was controlled by a DC servo delay stage (Thorlabs). For the source of the push pulse (~200 fs) a commercial optical parametric amplifier (OPA) OPHEUS ONE (Light Conversion) was used. For the 'slow' pump, 'slow' push experiments on AXT a second, OPA was also used as the pump pulse source. The push pulse was chopped synchronously with a chopper so as to generate the sequence of pulses shown in Fig 4a of the manuscript. Monitoring the push-probe signal ensures that any contribution due to excitation of the ground state by the push pulse is less than 1%. This second OPA was also used as the source in cryogenic pump-probe studies, two photon pump-probe spectroscopy experiments and in pump energy dependent measurements, further details of which are provided in the relevant supporting information.

Supporting Information and data availability

Sample characterisation of PDA and AXT, details of optical pulse compression, effect of temperature, pump fluence and pump energy on the pump-probe spectra of PDA and AXT, impulsive Raman spectroscopy of PDA, two-photon transient absorption spectroscopy of PDA, genetic algorithm decomposition of data, effect of push energy and time delay on pump-push-probe spectra of PDA and AXT.

The raw data and analysis codes associated with this publication are available at [url to be added in proof].

Author contributions

R.P., Q.G. and A.C. designed and performed the experiments and analysed the data. R.Y.S.C. developed the kinetic model and analysed the data. E.P.B. performed the X-Ray diffraction under the supervision of N.C.G.. T.B., L.L., M.S., R.S. and F.M. synthesised the PDA samples. A.J.M. provided the carotenoid samples and interpreted the data. R.P., A. R. and A.W.C conceived the project and wrote the manuscript with input from all authors.

Acknowledgements

We thank the Engineering and Physical Science Research Council (EPSRC, UK) and the Winton Program for the Physics of Sustainability for financial support. A.J.M. acknowledges support from EPSRC grant EP/M025330/1. R.P thanks A. Alvertis, C. Schnedermann, S. Dutton and J. Gorman (Cambridge) for useful discussions and advice.

Notes

The authors declare no competing financial interests.

During the preparation of this manuscript Michel Schott, Director of Research at CNRS, sadly passed away. This work is dedicated *in memoriam* to his outstanding career and lifetime contributions to the understanding of the photophysics of conjugated polymers.

For table of contents only

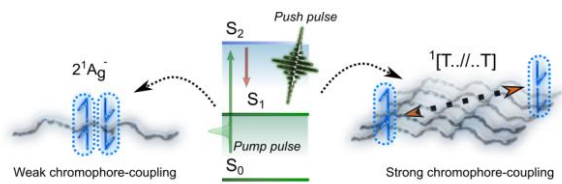